\documentclass[12pt,preprint]{aastex}

\begin{document}

\title{Contribution of GRB Emission to the GeV Extragalactic Diffuse 
Gamma-Ray Flux}

\author{S. Casanova and B. L. Dingus}
\affil{Los Alamos National Laboratory,
    Los Alamos, NM 87545}

\and

\author{Bing Zhang}
\affil{Physics Department, University of Nevada Las Vegas, NV 89154}

\begin{abstract}

TeV gamma rays emitted by GRBs are converted into electron-positron
pairs via interactions with the extragalactic infrared radiation
fields. In turn the pairs produced, whose trajectories are randomized
by magnetic fields, will inverse Compton scatter off the cosmic
microwave background photons. The beamed TeV gamma ray flux from GRBs
is thus transformed into a GeV isotropic gamma ray flux, which
contributes to the total extragalactic gamma-ray background emission.
Assuming a model for the extragalactic radiation fields, for the GRB
redshift distribution and for the GRB luminosity function, we evaluate 
the contribution of the GRB prompt and scattered emissions to the 
measured extragalactic gamma-ray flux. To estimate this contribution 
we optimistically require that the energy flux at TeV energies is 
about 10 times stronger than the energy flux at MeV energies. The
resulting gamma-ray diffuse background is only a small fraction of
what is observed, allowing blazars and other sources to give the
dominant contribution.
\end{abstract}


\section{Introduction}

The nature of the extragalactic gamma ray background emission has been
a topic of great interest since EGRET collaboration evaluated its
spectrum in the range from 30 MeV to 100 GeV \citep{Sreekumar:1997un}. The 
diffuse emission coming from beyond the galaxy was determined by 
subtracting the contributions of resolved point sources, the 
diffuse galactic emission, and the instrumental background from 
the gamma-ray intensities observed by EGRET. The emission is found to be a 
power law in energy
\begin{equation}
\frac{dN_{\gamma}}{ dA \, dt  \, d\Omega \, dE} = (7.32\pm0.34) \times
{10}^{-6} \quad {(\frac{E}{0.451 {\rm GeV}})} 
^{-2.10 \pm 0.03}  \, {\rm {cm}^{-2} {s}^{-1} {sr}^{-1} {GeV}^{-1}} 
\end{equation}
and is highly isotropic on the sky. \citet{Mukherjee:1999it} suggested 
that blazars can explain up to 25 per cent of the extragalactic 
emission. The hypothesis that the flux is predominantly due to blazars 
seems to be reinforced by a new evaluation of the extragalactic emission 
\citep{Strong:2004ry} which is slightly lower and steeper than that found by 
\citet{Sreekumar:1997un}. The result is not consistent with a power law 
and shows some positive curvature which the authors relate to an origin 
in blazars emission. \citet{Kneiske:2004rf} recently suggested that 
up to 85 per cent of the extragalactic emission could arise from blazars. 
According to \citet{Stecker:1996, Stecker:2001} blazars can account 
for the entire 
extragalactic $\gamma$-ray background observed by EGRET. Above 100 
MeV normal galaxies contribute from 3 to 10 per cent of the observed 
diffuse extragalactic flux. However from the analysis of \citep{Erlykin:1995} 
the spectrum of normal galaxies seems to differ from the extragalactic 
diffuse emission spectrum. \citet{Dar:1995} have suggested that the 
extragalactic diffuse arises from cosmic rays interacting with 
intergalactic gas. This contribution seems to disagree with the 
diffuse gamma ray spectrum according to \citet{Stecker:1996}. 
\cite{Loeb:2000} suggested fossil radiation from shock accelerated cosmic 
rays during structure formation as possible contribution. 
\citet{Stawarz:2006} have estimated that inverse Compton scattering 
of starlight photon fields by the ultrarelativistic electrons in 
kiloparsec-scale jets in FR I radio galaxies contributes about one 
percent to the EGRET extragalactic flux. \cite{Chi:1989} and \cite{Wdowczyk} 
have shown that upscattering of CMB to $\gamma$-ray energies 
by cosmic ray electrons and protons does not provide a 
sinificant contribution to the diffuse extragalactic emission. 

However, the question about the origin of the extragalactic 
emission is still open and can possibly be solved by admitting 
that different sources contribute to it. In fact all unresolved 
discrete sources outside the Galaxy contribute to the extragalactic 
background emission; the problem is clearly to understand how 
big the different contributions are. As already pointed out by 
\cite{Hartmann:2003}, who considered GRBs as a source for the 
diffuse gamma-ray emission at MeV energies, prompt and delayed emissions 
from GRBs should also contribute to the diffuse extragalactic emission, 
especially if some GRBs emit photons in the GeV-TeV energy. In fact, 
outside the GRB source, due to interactions with cosmic infra-red 
background photons, most of the high energy GRB photons produce high-energy 
electron-positron pairs. The pairs inverse Compton scatter off CMB 
photons and produce secondary photons, which in turn interact with IR 
photons and generate other pairs. Multiple inverse Compton scatterings 
occur until the energy of the secondary photons is no longer large 
enough to trigger pair production with the IR photons. When the energy 
of the scattered photons is not sufficient to produce a subsequent 
pair, the photons travel to the Earth without undergoing further 
absorption. After multiple pair-production and inverse Compton 
processes, the initial energy of the TeV photons escaping the GRBs 
will have been shifted to MeV-GeV energies 
\citep{Coppi:1997,Plaga:1995,Dai:2002gx,Wang:2004,Razzaque:2004cx}. The 
data on extragalactic diffuse emission in the MeV-GeV energy range are thus 
useful to put constraints on high energy emission from gamma ray 
bursts at GeV-TeV energy.

In the following we assume a flat cosmological model with normalized
Hubble constant $h=0.71$, with $\Omega_m=0.3$ and
$\Omega_{\Lambda}=0.7$.  Cosmological distances and volumes are
calculated following \cite{Hogg:1999ad}.

\section{GRB model}

Gamma ray bursts are cosmological spots of short, intense and narrowly
beamed $\gamma$-ray emission, whose observed isotropic luminosities,
$L_{iso}$, are in the range of ${10}^{51}-{10}^{52}$ erg/s. The huge
energy released by GRBs is almost certainly produced by
ultra-relativistic flows, whose bulk Lorentz factors $\Gamma_b$ may
vary in the range of 100 and 1000, with a typical value of 300
\citep{Baring:1997,Lithwick:2000kh,Zhang:2006}. The 
widely accepted fireball internal-external shock model for GRBs
envisages that high energy protons and electrons are accelerated in
shocks by interaction with magnetic fields. The process of Fermi
acceleration leads to electron and proton power-law spectra of index
between -2 and -3. Electrons and protons cool through synchrotron
emission. Since the electron cooling time is shorter than the proton
cooling time, as a first approximation, the GRB prompt spectrum is
believed to be dominated by electron synchrotron emission only. The
observed prompt spectrum arising from electron synchrotron emission is
a Band function \citep{Band:1993eg}, proportional to $E^{\alpha}$
below the synchrotron peak energy, $E_{pk}$, with $\alpha$ about -1
and to $E^{\beta}$ above 
the peak energy, with $\beta$ usually between -2 and -3. Below the
synchrotron self absorption energy, $E_{ssa}$, low energy photons are
absorbed by fireball electrons in a magnetic field by synchrotron
self-absorption. 
\clearpage
\begin{figure}[ht]                                        
\begin{center}
   \includegraphics[angle=0,width=12cm]{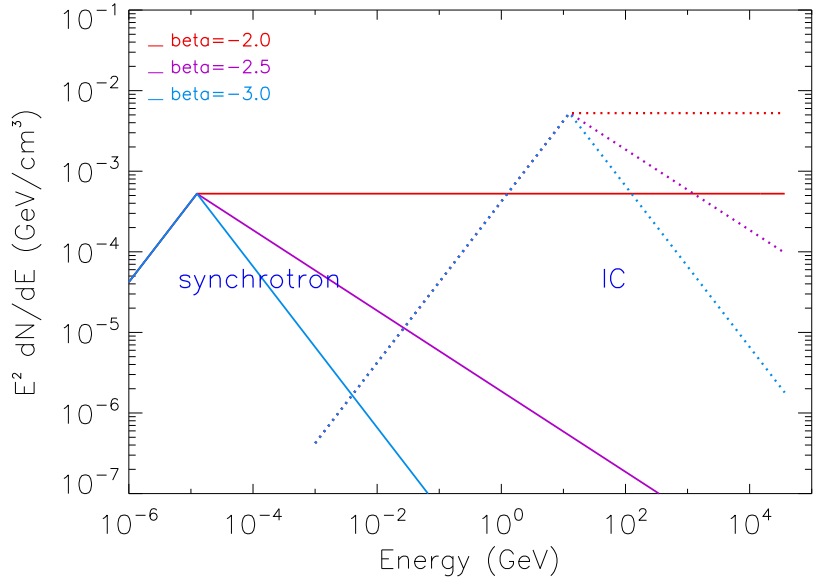}
\end{center}
\caption {Band spectrum and Inverse Compton spectrum for a GRB having
isotropic luminosity $10^{52}$ erg/s 
and located at $z=0.1$. The slope $\alpha=-1$ and ${\beta}$ varies
bewteen -2 and -3 for both synchrotron and IC spectra. The ratio $r$
between the synchrotron peak flux and the TeV emission peak flux is
assumed to be 10. \label{fig1}}
\end{figure} 
\clearpage
High energy (up to TeV) emission has been detected in some GRBs
\citep{Gonzales,Hurley:1994cf,Atkins:2002ws}. According to the
fireball model, TeV photons can be produced in both internal
\citep{Dai:2002gx,Razzaque:2004cx} and external
\citep{Zhang:2001,Boettcher:1998} shocks, either through electron
inverse Compton or proton synchrotron emission. \citet{Wang:2001} 
proposed that the cross inverse Compton scattering between the 
photons and electrons in forward and reverse shocks can also 
produce TeV photons. Tev photons can be also generated through
inverse Compton scattering of shock accelerated particles in
external shocks off a bath of photon that overlaps the shocked 
region. The photon bath could be either the prompt gamma-ray 
emission itself \citep{Beloborodov:2005,Fan:2005} or the
late-time X-ray flares \citep{Wang:2006}. For such distant 
sources as GRBs, TeV photons are mostly absorbed through pair 
production with cosmic background radiation (CBR), but some 
indications of TeV gamma rays from GRBs at low redshift were provided 
by experiments like Milagro \citep{Atkins}, Hegra \citep{Padilla} and 
Tibet \citep{Amenomori}. In our calculation TeV prompt energy photons 
are taken into account in the prompt spectrum without investigating 
the details of the way they are created. We model the GRB prompt TeV 
emission as an additional broken power law with low energy spectral 
index $\alpha=-1$ and high energy spectral index $\beta=-2$. The 
synchrotron peak energy $E_{pk}$ depends on the isotropic energy of 
the burst \citep{Amati:2002ny}. In order to match the typical observed 
$E_{pk}$ at 100 keV - 1 MeV, the typical random electron Lorentz 
factor $\gamma_e$ is about 300-1000. Since most high energy emission
components are related to inverse Compton (IC) scattering, we design 
our high energy component to mimic the IC in the internal shocks.
The typical separation 
between the synchrotron typical frequency and the IC typical frequency 
is $\gamma_e^2$, which varies between $10^5$ and $10^6$. So the break
energy of the additional TeV component is chosen to be around $10^5$ 
times the synchrotron peak $E_{pk}$ \citep{Zhang:2002jt}.
The relative energies contained in the synchrotron component and the 
IC component depend on the IC $Y$-parameter 
\citep{Panaitescu:2000,Sari:2001,Zhang:2001} 
\begin{equation} 
Y = \frac{L_{IC}}{L_{syn}} \sim \sqrt{(\epsilon_e/\epsilon_B)}
\end{equation} 
for $\epsilon_e \gg \epsilon_B$, where $\epsilon_e$ and $\epsilon_B$
are shock energy equipartition parameters for electrons and magnetic
fields, respectively. Because of a possible Klein-Nishina
limitation, this $Y$ parameter could not be too high. An optimistic
value would be around 10, corrsponding to $\epsilon_e \sim 100 
\epsilon_B$ (see Fig.~\ref{fig1}).
This requires that the GRB internal shocks are not very 
magnetized\footnote{Some evidence however suggests a magnetized 
central engine\citep{Zhang:2003,Fan:2002,Kumar:2003}.
In that case, the IC component is at most comparable to 
the synchrotron component.}.

GRB emission is likely narrowly beamed. 
The jet may be top-hat shaped with uniform energy distribution inside
the jet cone and a sharp drop off at the edge
\citep{Rhoads:1999wn,Sari:1999mr} 
or ``structured'' with angle-dependent energy density inside the jet
\citep{Zhang:2002,Rossi:2001pk}. 
The total amount of emission energy corrected for the beaming effect
is likely standard, 
about $ 1.3 \times {10}^{51}$ erg \citep{Frail:2001qp}. The average
beaming factor ranges in the literature from around 
75 \citep{Guetta:2004fc} to 500 \citep{Frail:2001qp}.

\begin{figure}[ht]                                        
\begin{center} 
\includegraphics[angle=0,width=11cm]{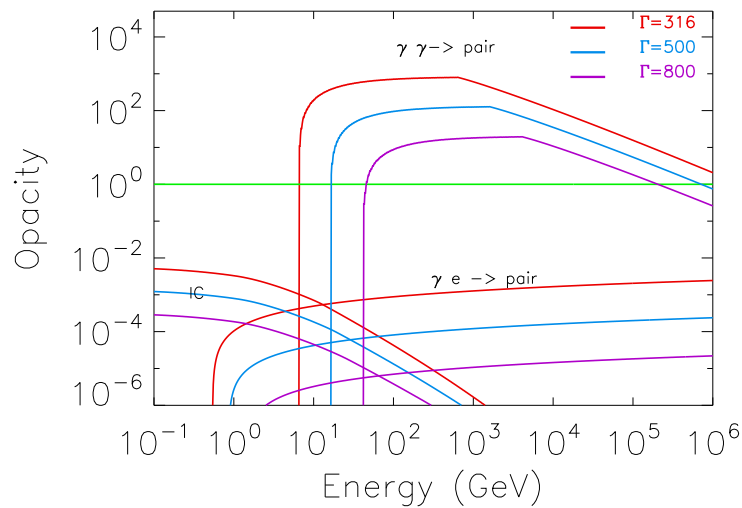}
\caption{Optical depths for processes involving $\gamma$-rays in the
GRB fireball. Inverse Compton and  
$\gamma-e$ pair production do not attenuate the photon spectrum
inside the GRB source. The only process attenuating the photon
spectrum is $\gamma$-$\gamma$ pair production. The attenuation is less
efficient for higher bulk Lorentz factors and for shorter variability
times. The case of a GRB having isotropic luminosity $10^{52}$ erg/s
are time variability 0.01 second for different bulk Lorentz factors is
plotted. \label{fig2}} 
\end{center}
\end{figure}

Photon-photon absorption through pair production inside the GRB source
region controls the range in energy and the amount of radiation
emitted. Our following treatments closely follow
\citet{Razzaque:2004cx}. In Fig.~\ref{fig2} we plot the optical depths 
corresponding to the different processes taking place inside the GRB
source region, $\gamma \gamma$ absorption, electron Compton scattering
and $e \, \gamma \rightarrow e^{\pm}$. The GRB isotropic luminosity is
$L_{iso}={10}^{52} \, {\rm erg/s}$ and the observed time variability is
$\delta t =0.01$ s. The time variability is an important parameter for
the description of gamma ray bursts. In fact the shock radius
\begin{equation}
r_{sh}= 2 \, c \, \delta t \, \Gamma_b^2
\end{equation}
is proportional to the time variability $\delta t$ and the peak volume 
number density of photons is defined as 
\begin{equation}
{n'}_{\gamma} = \frac{L_{iso}}{4 \, \pi \, {r_{sh}}^2 \, c \, {\Gamma}_b 
\, 
E_{pk}} \,.
\end{equation}  
As shown in Fig.~\ref{fig2}, $\gamma \gamma$ pair production within the 
GRB
attenuates the prompt spectrum, while electron Compton scattering and
$e \, \gamma \rightarrow e^{\pm}$ have optical depths less than 1 for
the typical parameter sets we are adopting.
The optical depth for $\gamma$-ray absorption through pair production 
is the integral above the synchrotron self absorption 
energy $E^{'}_{ssa}$ and below 
the Klein Nishima limit energy $E^{'}_{KN}
= {\Gamma}_b \, m_e \,{{\gamma}^{'}_{e}}_{max}$ 
where the maximum electron Lorentz factor is ${{\gamma}^{'}_{e}}_{KN}
=\frac{3 \, e}{{\sigma}_T \, B \, (1+Y)}$. \citep{Dai:2002gx}
\begin{equation}
\tau_{grb} = \frac{r_{sh}}{\Gamma_b} \, \int_{{E'}_{ssa}}^{{E'}_{max2}} \sigma_{pair} \,
\frac{d{N'}_{\gamma}}{d{E'}_{\gamma}} \, d{E'}_{\gamma} \,.
\label{eqn:tau}
\end{equation}
In Eq.~\ref{eqn:tau} the cross section for $\gamma$-$\gamma$ pair 
production $\sigma_{pair}$ is
\begin{equation}
\sigma_{pair}= \frac{3}{16} \, \sigma_{th}(1-\beta^2) \, (2 \, \beta
\, (\beta^2-2)+(3-\beta^4 \, log((1+\beta)/(1-\beta))) 
\end{equation}
with
\begin{equation}
\beta= \sqrt{1 - \frac{{(m \, c^2)}^2}{E_{\gamma} \, \epsilon_{cmb}}} \,.
\end{equation}
$\frac{d{N'}_{\gamma}}{d{E'}_{\gamma}}$ in Eq.~(\ref{eqn:tau}) is the
number density of photons per energy in the comoving frame arising
from synchrotron and higher energy processes.  
The total number density of photons is given as a superposition of
different synchrotron and 
possibly higher energy power law spectra 
\begin{equation}
\frac{d {N'}_{\gamma}}{d{E'}_{\gamma}}  =  \frac{d 
{{N_1}'}_{\gamma}}{d{E'}_{\gamma}} + 
\frac{d {{N_2}'}_{\gamma}}{d{E'}_{\gamma}} +  \frac{d 
{{N_3}'}_{\gamma}}{d{E'}_{\gamma}} + 
\frac{d {{N_4}'}_{\gamma}}{d{E'}_{\gamma}}
\end{equation}
where
\begin{eqnarray*}
\frac{d {{N_1}'}_{\gamma}}{d{E'}_{\gamma}} &=&
\frac{{n'}_{\gamma}}{{{E'}_{pk}}}\, 
(\frac{{{E'}_{\gamma}}}{{E'}_{pk}})^{\alpha} 
\quad \quad \mbox{for ${E'}_{ssa} < {E'}_\gamma < {E'}_{pk} $}  \\[2mm]
\frac{d {{N_2}'}_{\gamma}}{d{E'}_{\gamma}} &=&
\frac{{n'}_{\gamma}} {{{E'}_{pk}}} 
(\frac{{{E'}_{\gamma}}}{{E'}_{pk}})^{\beta} 
\quad \quad \mbox{for ${{E'}_\gamma}_{pk} < {E'}_{\gamma} < {E'}_{max}$} 
\\[2mm] 
\frac{d {{N_3}'}_{\gamma}}{d{E'}_{\gamma}} &=&
C \, \frac{{n'}_{\gamma}}{{{E'}_{pk2}}}\, (\frac{{{E'}_{\gamma}}}
{{E'}_{pk2}})^{\alpha} 
\quad \quad \mbox{for ${E'}_{min} < {E'}_\gamma < {E'}_{pk2} $}  \\[2mm]
\frac{d {{N_4}'}_{\gamma}}{d{E'}_{\gamma}} &=&
C \, \frac{{n'}_{\gamma}} {{{E'}_{pk2}}} 
(\frac{{{E'}_{\gamma}}}{{E'}_{pk2}})^{\beta} 
\quad \quad \mbox{for ${E'}_{\gamma} > {E'}_{pk2}$}  \,.
\end{eqnarray*}
\begin{equation}
\label{eqn:fluxcomoving}
\end{equation}
${E'}_{pk}$ and ${E'}_{pk2}$ are the synchrotron peak energy and the
higher emission peak energy, respectively. ${E'}_{ssa}$ is the
synchrotron self absorption energy, ${E'}_{max}$ is the synchrotron cut off 
corresponding to the maximum energy the electrons are 
accelerated to by the Fermi mechanism and ${E'}_{KN}$ the IC 
Klein-Nishima limit. ${E'}_{min}$ is the IC-boost of 
the synchrotron self-absorption frequency. The ratio $R$ of the high 
energy component energy flux, $({{E'}_{pk2}})^2 \, \frac{d{N'}_{\gamma}}{d{E'}_{\gamma}}$, 
and the synchrotron energy flux, $({{E'}_{pk}})^2 \, 
\frac{d {N'}_{\gamma}}{d{E'}_{\gamma}}$ at the peaks is given by 
\begin{equation}
R = C \frac{E_{pk2}}{E_{pk}},
\end{equation}
which is adopted as the most optimistic value $R\sim 10$ in our
calculations (see above for more discussion).
Here $\alpha$ and $\beta$ are free parameters, which we assume to be
$\alpha=-1$ and $\beta=-2$. 
\clearpage
\clearpage
Prompt photons are absorbed above the cutoff energy $E_{cutoff}$ 
\begin{equation}
E_{cutoff} = \frac{m_e^2 \,c^4 {\Gamma_b}^2}{2 \,E_{pk}} \, ,
\end{equation}
corresponding to the photon threshold energy for pair production with
lower energy photons. The prompt photons are not absorbed above the so
called thinning energy, the photon energy corresponding to optical
depth $\tau_{\gamma \gamma} = 1$
\begin{equation}
E_{thinning} = \frac{3 \Lambda L_{iso} \sigma_{TH} m_{e}^2 }{64 \pi
{\Gamma}^2 \delta t \, {E_{\gamma \, ssa}}^2} 
\end{equation}
where $\Lambda= log[2 \, {(2 {E_{\gamma \, ssa}}
E_{\gamma})}^{1/2}]/(m_e \, \Gamma_b)$. For high enough bulk Lorentz
factors and for short enough time variabilities, GRBs are optically
thin at all energies. TeV energy photons emitted by some processes
within the GRB will thus be able to leave the source and be observed
by experiments like Milagro.

\section{Photon absorption off IR photon fields}

\begin{figure}[ht]
\begin{center}
\includegraphics[angle=0,width=11cm]{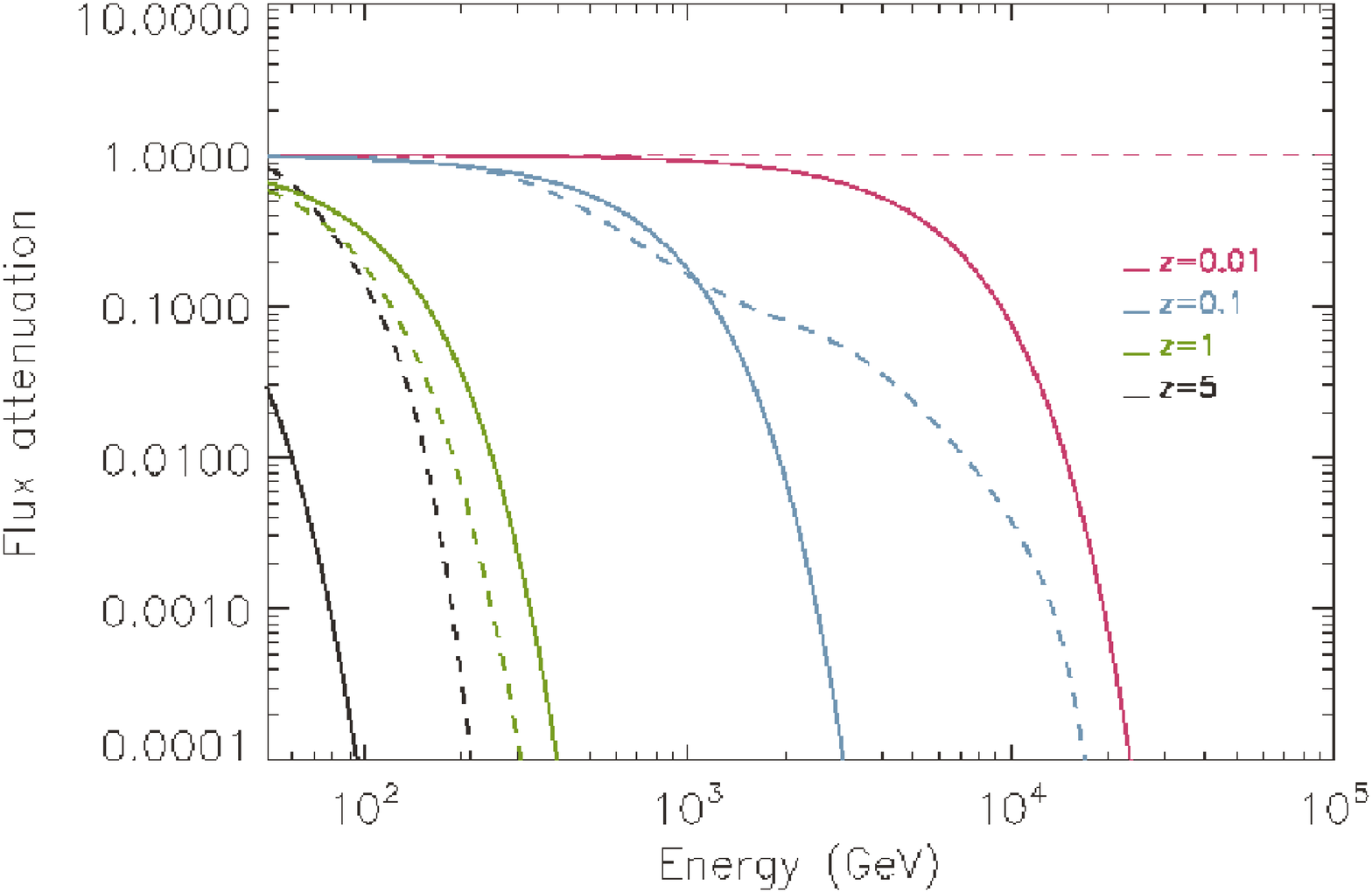}
\caption {Attenuation of $\gamma$-ray spectra through $\gamma \gamma$
absorption at different redshifts. The continous lines show the parametrization 
of Eq.~\ref{kneiske} and the dashed lines the best fit model 
of \citet{Kneiske:2003tx}. \label{fig3}}
\end{center} 
\end{figure}

High energy gamma rays are attenuated also when travelling to us
because they form pairs in collisions with low energy photons from the
meta-galactic radiation field. Following \citet{Kneiske:2003tx}, the
optical depth of gamma rays depends on 
the redshift $z$ and the gamma ray energy $E_\gamma$ is parametrized by
\begin{equation}
{\tau_{bkg}}_{\gamma \gamma}(E_{\gamma},z) = z^{1.33} \,
{(E_{\gamma}/E_0)}^{3/2} 
\label{kneiske}
\end{equation}
where $E_0=90$ GeV. The attenuation rate versus energy is plotted for
different redshifts in Fig.~\ref{fig3}. The continous lines show 
the parametrization given in Eq.~\ref{kneiske}, whereas the dashed 
lines represent the best fit model by \citet{Kneiske:2003tx}. \citet{Stecker:2006}'s 
recent calculations of 
intergalactic gamma-ray absorption making use of new Spitzer and GALEX data diverge from 
those of \citet{Kneiske:2004rf} at the higher redshifts due to the recent discovery 
that active star formation was taking place in young galaxies at redshifts out to beyond 6. 
This gives larger optical depths at the higher redshifts than previously thought.
\clearpage
\clearpage
\section{Prompt and scattered emission}

Throughout the rest of the paper we will assume that both the synchrotron and the 
higher energy GRB spectrum have spectral indices $\alpha=-1$ and $\beta=-2$. 
Also we will consider GRB having time variability of 1 seconds and duration of 
20 seconds and a bulk Lorentz factor of 316.
 
Inside and outside the GRB, the spectrum is assumed to be only
attenuated by $\gamma \gamma$ reactions and the flux which leaves the
source and is observed at a luminosity distance $D_L(z)$ is
\begin{equation}
\frac{{d N_{\gamma}}_{prompt}}{dE_{\gamma} \, d t \, d A} (E_{\gamma},
z,L_{iso}) =
 \frac{L_{iso}}{{4 \, \pi \,{{{D}^2_L(z)}}}\, E_{pk} \, } \, f(E_\gamma) 
\,e^{-\tau_{grb \,\gamma \gamma}(E_{\gamma})}  
\, \, e^{-{\tau_{bkg \,\gamma \gamma}(E_\gamma,z)}}
\label{eqn:fluxemitted}
\end{equation}
where
\begin{equation} 
f(E_\gamma) = \frac{1}{{n'}_{\gamma} \, {\Gamma}_b} \, \frac{d
{N^{'}}_{\gamma}}{{dE^{'}}_{\gamma}} 
\end{equation}
and $\tau_{grb \, \gamma \gamma}$ and ${\tau_{bkg \, \gamma \gamma}}$
are the optical depths for pair production inside and outside the
source.
\clearpage
\begin{figure}[ht]
\begin{center}
\includegraphics[angle=0,width=12cm]{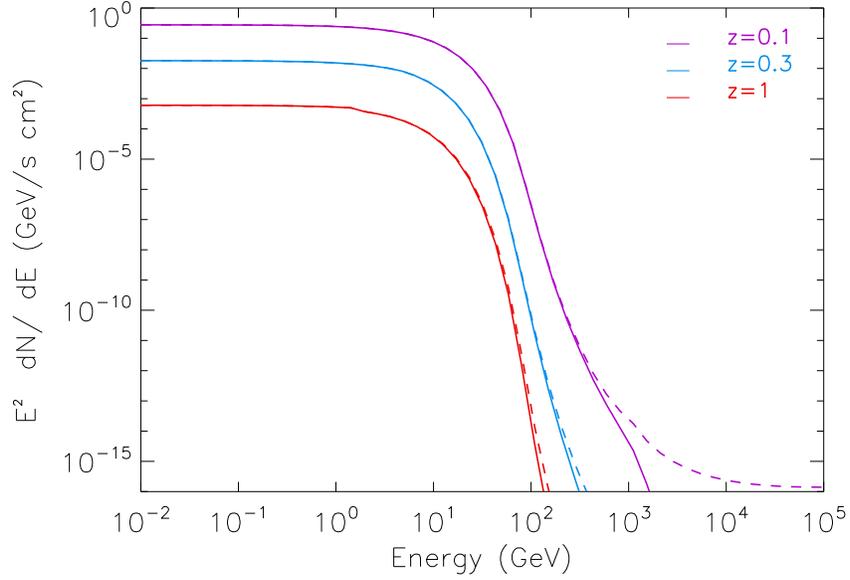}
\caption {Synchrotron and higher energy prompt flux from one GRB
having isotropic luminosity $10^{52}$ erg/s, bulk Lorentz factor 316
and time variability 1 second.  The slopes $\alpha=-1$ and $\beta=-2$
for both synchrotron and IC spectra. The ratio between the synchrotron
peak flux and the TeV emission peak flux is assumed to be 10.  The solid 
lines represented the fluxes attenuated by $\gamma \gamma$ absorption inside and outside
the sources which are located at redshifts 0.1, 0.3 and 1. 
The dashed lines indicate the fluxes after attenuation inside 
the GRB. \label{fig4}} \end{center} 
\end{figure}   
\clearpage
Outside the GRB, due to $\gamma \gamma$ interactions with cosmic
infra-red background photons (IR) and the cosmic microwave background
photons (CMB), most of the high energy photons produce high-energy
$e^{\pm}$ pairs. We assume that each lepton of the pair shares half of
the energy of the initial GRB photon, $E_e = {E_\gamma}/{2}$. The
electron-positron flux from the synchrotron and the IC fluxes from one
GRB at a distance $D_L(z)$ in the observer's frame is
\begin{equation}
{\left(\frac{ {d N_e} }{dA \, dE_e \, dt} (E_e,z,L_{iso})\right)}_{grb} =   
2 
\frac{L_{iso}}{{4 \, \pi \,{{D}^2_L(z)}}} \, \, \frac{f(E_\gamma)}{E_{pk}} 
\, \, 
e^{-\tau_{grb \,\gamma \gamma}(2  \, E_{e})} 
\, [ 1 - e^{-{\tau_{bkg \,\gamma \gamma}(2  \,E_e,z)}}]   
\label{eqn:pairs}
\end{equation}
The pair flux will be enriched as long as the burst lasts. The time
integrated electron-positron fluxes are given by
\begin{equation}
{(\frac{dN_{e}}{dA \,d\,E_e})}_{grb}= \int_{0}^{T} \, dt \,
{(\frac{dN_e}{dA \, d\,E_e \, dt})}_{grb} \,.
\end{equation}
The pairs will inverse Compton scatter off CMB and IR photons and
produce secondary photons, which will in turn interact with IR photons
and generate other pairs. Multiple inverse Compton scatterings will
happen until the energy of the secondary photons is no longer
sufficient to trigger a pair production with the IR photons. When the
energy of the scattered photons is insufficient to produce a
subsequent pair, the photons will travel to the detector without
undergoing further absorption. When simulating the series of inverse
Compton scatterings we assume that all interactions happen very close
to the source, at the redshift of the source itself. We also assume
that the magnetic field is stronger than $10^{-16}$ Gauss in order for
the flux to be isotropically radiated.  The number of secondary
photons per unit volume per photon energy interval created in the
vicinity of a GRB through one inverse Compton scattering of the pairs
in Eq.(\ref{eqn:pairs}) off cosmic of the background radiation photons
is \cite{Gaisser}
\begin{eqnarray*}
&& {{(q_{IC}(E_\gamma,z,L_{iso}))}_{grb}} =({\frac{d \,N_{\gamma} }{d \, 
E_{\gamma} \, d V}})_{grb}  \\[2mm]
&=&  \int d {E_e} \int 
d {\epsilon}_{cmb} 
\frac{d {\sigma_{IC}}} {d E_{\gamma}}( E_\gamma, {E_e} \,\epsilon_{cmb}) 
\, u_{cmb}( {\epsilon}_{cmb},z) \,  
({\frac{d N_e}{dA \, d{E_e}}})_{grb}
\end{eqnarray*}
\begin{equation}
\label{inverse_integral1}
\end{equation}
where $\frac{d \sigma_{IC}(E_\gamma,E_e,\epsilon_{cmb})}{d
E_{\gamma}}$ can be either the differential Thomson cross section (for
low energy pairs) or the differential Klein-Nishina formula (for high
energy pairs) \citep{Schlickeiser}, $u_{cmb}( {\epsilon}_{cmb},z)$ is
the density of CBR photons per unit volume per photon energy interval
at redshift $z$ \citep{Peebles}.  If inverse Compton scattered photons
have a high enough energy, they will be absorbed by the IR photons.

In order to evaluate the contribution from scattered $\gamma$-ray
emission from all GRBs we input the observed GRB redshift distribution
\citep{Guetta:2004fc,Firmani:2004fn}
\begin{equation}
n_{GRB}(z) = \frac{R_{GRB}(z)}{1+z} \,  \frac{d V}{dz} \, \int \, d  \, 
log \, (L_{iso})  \, \Phi_0(L_{iso}) \,.
\label{dist_guetta}
\end{equation} 
In Eq.(\ref{dist_guetta}), $\Phi_0(L_{iso})$ is the luminosity
function, defined as the comoving space density of GRBs in the
interval $log(L_{iso})$ and  $log(L_{iso}) + d \,
log(L_{iso})$. Following \citet{Schmidt:1999iw}, we derive the  
luminosity function $\Phi_0(L_{iso})$ by assuming 
a broken power law with lower and upper limit, ${L}_{lower}=
\frac{{L_{iso}}^{*}} {30}$ and  ${L}_{upper}=10 \, {L_{iso}}^{*}$, 
respectively, where ${L_{iso}}^{*}= 4.4 \,{10}^{51}$ erg/s, 
\begin{eqnarray}
\Phi_0(L_{iso}) = \left\{ \begin{array}{c@{\hspace{12mm}}l}
a \, {(\frac{L_{iso}}{L_{iso}^{*}})}^{\gamma_1}  & \mbox{for $L_{lower} < 
{L_{iso}}^{*} < L_{upper}$}  \\
a \, {(\frac{L_{iso}}{{L_{iso}}^{*}})}^{\gamma_2}  & \mbox{for $L_{lower} 
< {L_{iso}}^{*} < L_{upper}$}  \\
\end{array} \right. 
\label{luminosity_function}
\end{eqnarray}
The normalization constant $a$ 
\begin{equation}
a = \frac{1}{\frac{1}{\gamma_1} \, (1-\frac{1}{\delta_1^{\gamma_1}}) +
 \frac{1}{\gamma_2} \, (-1+\delta_2^{\gamma_2})}
\end{equation}
is obtained by imposing the integral of $\Phi_0(L_{iso})$ to give
unity. We assume $\gamma_1=-0.1$, $\gamma_2=-2$, $\delta_1=30$ and
$\delta_2=10$. 

The integration over 
$L_{iso}$ is performed over the interval indicated in 
\cite{Guetta:2004fc}.
In Eq.(\ref{dist_guetta}) the redshift distribution of GRBs $R_{GRB}(z)$ 
is   
\begin{equation}
R_{GRB}(z) = \frac{ 23 \, e^{3.4 z} \, \rho_{GRB} \, G(z, \Omega_m, 
\Omega_{\Lambda})}{ 22 + e^{3.4 z} }
\end{equation} 
where 
\begin{equation}
G(z, \Omega_m, \Omega_{\Lambda}) =  \frac{\sqrt{\Omega_m \, {(1+z)}^3 + 
\Omega_k \, {(1+z)}^2 + \Omega_{\Lambda}}}{
{(1+z)^{3/2}}}
\end{equation} 
and $\rho_{GRB} = 0.44 \,{Gpc}^{-3} \, {yr}^{-1}$ is the observed rate
of GRB per differential co-moving volume.
\begin{equation}
\frac{d V}{dz} = \frac{d_h \,(1+z)^2 {d_a}^2 \,d\,\Omega}{\sqrt{\Omega_m 
\,{(1+z)}^3 + \Omega_k \, {(1+z)}^2 + \Omega_{\Lambda}}} \,. 
\label{comoving}
\end{equation}
where $d_a$ is the angular diameter distance and $d_h$ is the Hubble
distance. 

The secondary source function in Eq.(\ref{inverse_integral1}),
iterated over multiple scatterings, is integrated over the  
comoving line of sight distance $dl$ 
\begin{equation}
dl = d_{h}\frac{dz}{\sqrt{\Omega_m \, {(1+z)}^3 + \Omega_k \, {(1+z)}^2 + 
\Omega_{\Lambda}}} 
\label{lineofsight}
\end{equation}
and weighted over the number of GRBs $n_{GRB}$ per unit time to obtain the total
counting rate of the diffuse background flux
\begin{equation}
F_{scattered} = \frac{{d N_{\gamma}}_{scattered}}{d A \, dE_{\gamma} 
\,dt}  = \int d\, l  \, \int dz \,\, {(q_{IC}(E_\gamma,z,L_{iso}))}_{grb} 
\, n_{GRB}(z) \, e^{-{\tau_{bkg}}_{\gamma \gamma}(E_\gamma,z)}
\label{delayed_partial}
\end{equation}
where the integration over the redshift is performed from redshift
0.01, corresponding to about 40 Mpc, to redshift 10. As we mentioned
earlier if we assume that the intergalactic magnetic field is stronger
than $10^{-16}$ Gauss the beamed emission from GRBs is isotropically
radiated. So the flux contribution for each burst is lower by the
beaming solid angle divided by $4 \,\pi$. However there are more GRBs,
increased by $4\,\pi$ divided by the beaming solid angle, which
contribute to the diffuse background. So the two beaming factor
corrections cancel out and the unknown beaming fraction does not
influence Eq.(\ref{delayed_partial}).

The total flux from the prompt GRB emission is given by the following 
integral
\begin{equation}
F_{prompt} =  \frac{{d N_{\gamma}}_{prompt}}{d A \, dE_{\gamma} \,dt} =  
\, \int d z \, 
{(\frac{{d \,N_{\gamma}}_{prompt} } {d \, E_{\gamma} \, d A })}_{grb} \, 
n_{GRB}(z) \, \,e^{-{\tau_{bkg}}_{\gamma \gamma}(E_\gamma,z)}   
\label{prompt_integral}
\end{equation}
where 
\begin{equation}
{(\frac{{d \,N_{\gamma}}_{prompt} }{d \, E_{\gamma} \, d A })}_{grb}=  
\int_{0}^{T} dt \, {(\frac{{d \,N_{\gamma}}_{prompt} }{d \, E_{\gamma} d t 
\, d A })}_{grb} 
\end{equation}
is the differential gamma ray prompt flux from each GRB in
Eq.(\ref{eqn:fluxemitted}), measured in ${\rm {GeV}^{-1} {cm}^{-2}}$,
$n_{GRB}(z)$ is the redshift dependent number of GRBs per unit time
given in Eq.(\ref{dist_guetta}).

In Fig.~\ref{fig5} we plot the sum of the prompt and scattered GRB 
emissions
assuming an average value of the bulk Lorentz factor equal to 316. The
average time variability of the bursts is assumed to be 1 second and
the duration 20 seconds. The input energy flux at the high energy peak
$E_{pk2}$ is assumed 10 times stronger than that at the synchrotron peak
$E_{pk}$, i.e. $R=10$ (see Fig.~\ref{fig1}).

For higher average bulk Lorentz factors $\Gamma_b$, the pair 
production inside the GRB
source region is less efficient and therefore the $\gamma$-ray flux
escaping the GRBs will be higher and both the prompt and the scattered
fluxes in Fig.~\ref{fig5} will be higher. If the average time variability
$\delta t$, which is related to the dimensions of the GRB shock
radius, is lower, then the number of photons emitted through electron
synchrotron emission inside the GRB source region is higher and
consequently we expect higher prompt and scattered fluxes. The second
peak energy $E_{pk2}$ is the energy at which we expect the maximum TeV
energy flux and, as pointed out, is constrained by the observed
synchrotron spectrum to be between ${10}^5$ and ${10}^6$ times the
synchrotron peak energy $E_{pk}$. If we consider a lower value of
$E_{pk2}$, we have more photons at lower energies which scatter off
the CBM and IR photons and therefore the cascade processes started by
the GRB photons off the interstellar radiation fields will be
interrupted sooner and will produce lower scattered photon fluxes. A
lower second peak energy $E_{pk2}$ might also imply a more efficient
pair production absorption inside the GRB source region, as shown in
Fig.~\ref{fig2}. Finally assuming a shorter average time duration $T$ for 
the
GRBs less photons will be injected by the GRBs and the scattered flux
will be lower. 

As a back of the envelope calculation, by assuming that
the average amount of energy released by each GRB is of the order of
${10}^{51} \, erg $, the GRB rate is one per day and the average
redshift is 1, the total flux emitted by all GRBs is roughly of the
order of $5 \times {10}^{-9} \, GeV \, cm^{-2} \, s^{-1} \,
{sr}^{-1}$. This is roughly consistent with our more rigorous
calculations. 

\section{Conclusions and discussion}

Current limits for the contribution of blazars and other sources to the extragalactic
diffuse emission indicate that between 25 and 50 per cent of the
extragalactic diffuse emission is not explained yet and, in principle,
can be due to any other unresolved source outside the
Galaxy \citep{Sreekumar:1997un, Mukherjee:1999it, Hartmann:2003, 
Strong:2004ry, Kneiske:2004rf, Stecker:1996, Stecker:2001, 
Dar:1995, Loeb:2000, Stawarz:2006}. The prompt 
and scattered emissions from GRBs
should contribute to the diffuse extragalactic emission, too. In fact,
outside the GRB source, due to interactions with cosmic infra-red
background photons, most of the high energy GRB photons produce produce
high-energy electron-positron pairs. The pairs inverse Compton scatter
off CMB photons and produce electromagnetic cascades. In this way the
beamed GeV-TeV GRB emission is re-processed and converted into an
isotropic MeV energy emission, which contributes to the extragalactic
diffuse emission. 

\begin{figure}[ht]
\begin{center} 
\includegraphics[angle=0,width=12cm]{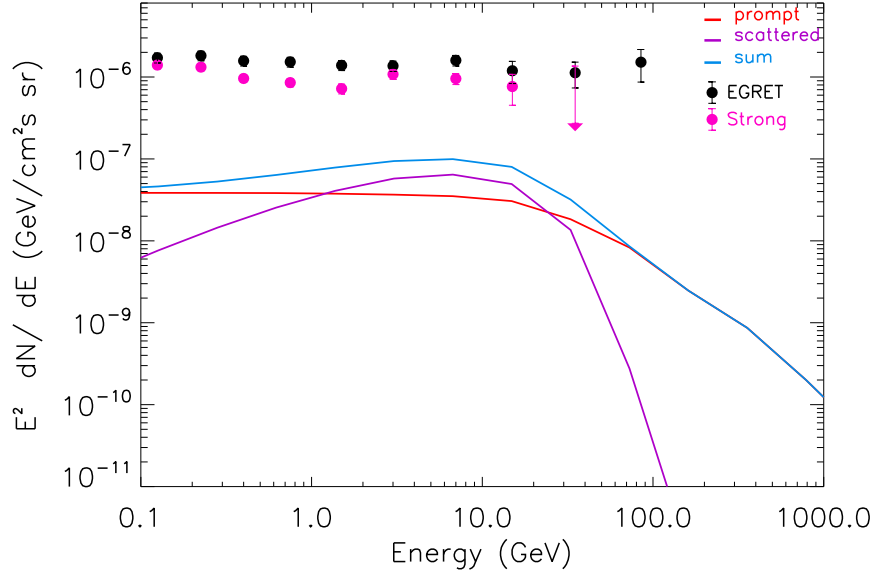}
\caption {GRB Total Emission from bursts having average bulk Lorentz
factors 316, time variability 1 second and duration 20 seconds.  The
slopes $\alpha=-1$ and $\beta=-2$ for both synchrotron and IC
spectra. The ratio between the synchrotron peak flux and the TeV
emission peak flux is assumed to be 10. The flux is attenuated by
$\gamma \gamma$ absorption inside and outside the sources. \label{fig5}}
\end{center} 
\end{figure}

We have modeled the emission from GRBs as due a low energy
synchrotron spectrum and a higher energy IC spectrum. The IC spectrum extends 
up to Klein-Nishima limit. Also we 
have assumed the higher energy IC emission from GRBs to be 10 times
stronger than the lower energy standard synchrotron emission. This 
assumption is equivalent to requiring that the GRB inverse Compton 
Y-paramater is equal to 10. Using the BATSE peak flux distribution 
\citet{Guetta:2004fc} derived the GRB luminosity 
function assumed for our estimates. The luminosity function 
can be approximated by a broken power law with a break peak luminosity of 
$4.4 \times {10}^{51} {\mathrm {ergs}} {\mathrm {s^{-1}}}$, a typical jet 
angle of 0.12 rad, and a local GRB rate of $0.44 h^3 {\mathrm{{Gpc}^{-3}}} 
{\mathrm {{yr}^{-1}}}$. Using the redshift and luminosity distributions 
derived by \citet{Guetta:2004fc} we have summed up 
the prompt and scattered emissions from all GRBs in the universe. The
sum of the prompt and scattered emission from all GRBs is shown in
Fig.~\ref{fig5} for a particular choice of GRB parameters. From our 
optimistic model the $\gamma$-ray emission from prompt and 
scattered GRB emissions can provide only a small fraction of 
the extragalactic diffuse emission at EGRET energies and cannot 
therefore explain the missing flux.
From Fig.~\ref{fig5} the sum of prompt and scattered 
emission from synchrotron and 10 times stronger IC emissions from all GRBs 
is less than the part of the extragalactic diffuse emission not explained 
by blazars, admitting that the blazars explain between 25 and 50 percent 
of the isotropic extragalactic emission. 
Our 
optimistic 
GRB model does not overproduce the extragalactic $\gamma$-ray 
background. We can thus conclude that TeV energy prompt fluxes 
from GRBs can be at least 10 times bigger than the synchrotron 
flux without violating the limit imposed by the extragalactic 
diffuse emission. In allowing TeV energy fluxes from GRBs of comparable 
or even stronger intensity to the MeV fluxes, our result is consistent with 
those models \citep{Peer} which predict fluences at TeV energies 
similar to those at MeV energies, the observations of high energy 
emission from GRB 970417a with Milagrito \citep{Atkins:2002ws} and 
the analysis of GRB 941017 done by \citet{Gonzales}, which pointed out 
the existence of a second flux from GRB 941017 that cannot be explained as 
synchrotron emission.  Experiments like Milagro and future missions 
like GLAST or mini-HAWC should therefore devote part of their efforts 
to investigate VHE emissions from GRBs.

We note that the recent Swift detection of GRB 060218
\citep{Campana:2006} suggests that there might be a low-luminosity
population of GRBs with a much higher event rate
\citep{Liang:2006}. On the other hand, these bursts have low
luminosities and tend to be X-ray flashes \citep{Campana:2006}, which
would compensate their high event rate. Future analyses are needed to
reveal the contribution from this category of GRBs.

\clearpage
\clearpage
 
\begin{acknowledgements}
The authors would like to thank Pablo Saz Parkinson.
\end{acknowledgements}


\end{document}